# Demonstration of Electron-Mediated Voltage-Controlled Exchange Coupling in Perpendicular Magnetic Tunnel Junctions

Qi Jia,[1] Yu-Chia Chen, [1] Delin Zhang, [1] Yang Lv, [1] Shuang Liang, [2] Onri Jay Benally, [1] Yifei Yang, [1] Brahmdutta Dixit, [1] Deyuan Lyu, [1] Brandon Zink[1] and Jian-Ping Wang*[1]

[1]Department of Electrical and Computer Engineering, University of Minnesota, 200 Union St. SE, Minneapolis, MN, 55455, USA
[2]Department of Chemical Engineering and Materials Science, University of Minnesota, 421 Washington Ave. SE, Minneapolis, MN, 55455, USA

**ABSTRACT**. Electron-mediated voltage control of exchange coupling (EM-VCEC) has been proposed as a mechanism for magnetization switching via modulation of spin-dependent electron reflection. However, its experimental verification has been challenging due to the coexistence of slower, voltage-induced ionic effects. Here, we fabricate magnetic tunnel junction (MTJ) devices that enable nanosecond timescale voltage application. Our results reveal rapid exchange coupling modulation on the nanosecond timescale, consistent with an electronic origin. The observed enhancement and saturation at low temperatures further rule out ionic migration, conclusively confirming the electronic nature of the mechanism. These results establish EM-VCEC as a viable mechanism for fast and energy-efficient voltage-driven magnetic switching.

Voltage-induced control of magnetism not only deepens the understanding of spin–charge interactions but also offers a non-dissipative magnetization manipulation [1–3]. The discovery of voltage-controlled magnetic anisotropy (VCMA) enables electric-field tuning of ferromagnetism via modulation of interfacial orbital occupancy, laying the foundation for low-power spintronics memory devices [4–12]. However, VCMA preserves time-reversal symmetry and thus prohibits deterministic bipolar switching. In contrast, VCEC breaks the time-reversal symmetry through mechanisms such as the magneto-ionic effect [13–17], strain-mediated coupling [18–20], and spin-dependent electron reflection [21–27], enabling bipolar switching. The first two rely on ionic migration or piezoelectric deformation and are limited by slow response times (milliseconds to microseconds). In contrast, the electronic mechanism enables nanosecond-scale switching, offering a fast and robust route toward high-speed spintronic applications.

VCEC induced by spin-dependent electron reflection, hereby referred to as EM-VCEC, was theoretically proposed by You [28] and Fechner [29] based on Bruno's model [30]. Subsequent experiments in MgO-based magnetic tunnel junctions reported voltage-controlled effective fields and magnetization switching, attributed to EM-VCEC [21–23,25–27]. However, these studies were performed under quasi-equilibrium conditions at room temperature, where concurrent ionic effects may have contributed, leaving the purely electronic origin of the EM-VCEC experimentally unconfirmed.

To isolate the electronic contribution to VCEC, we design a perpendicular magnetic tunnel junction (pMTJ) with a free layer coupled via the Ruderman–Kittel–Kasuya–Yosida (RKKY) interaction. The stack is fabricated into nanopillars with custom-designed electrodes to support nanosecond-scale voltage application, thereby excluding slower, non-electronic effects. We observe a rapid modulation of the effective exchange field, sufficient to induce switching, demonstrating the fast response of EM-VCEC. Moreover, the enhancement and eventual saturation of VCEC strength under DC bias at low temperatures exclude ionic migration, further supporting its electronic origin.

We deposited a multilayer stack on a $SiO_2$ substrate with the structure (thickness in nm): Substrate / Ta (5) / Pd (3) / Co (0.3) / Pd (0.7) / Co (0.3) / Ru (0.6) / Ta (0.3) / CoFeB (1.0) / MgO (2.0) / CoFeB (1.3) / Ta (0.7) / $[Pd (0.7)/Co (0.3)]_4$ / Pd (5). As illustrated in Fig. 1b, the device comprises four key components: [1] a fixed layer consisting of CoFeB (1.3) / Ta (0.7) / [Pd (0.7) / Co $(0.3)]_4$, providing a stable reference magnetization; [2] a free layer, CoFeB (1.0), located beneath the MgO barrier and subject to voltage control; [3] a coupling layer, [Co (0.3) / Pd (0.7) / Co(0.3)], which generates the exchange field; [4] a spacer, Ru (0.6) / Ta (0.3), mediating the coupling between the free and coupling layers. A relatively thick MgO barrier (2 nm) was selected to suppress current-induced effects while allowing the application of higher voltages.

After rapid thermal annealing, the PMA of the stack was confirmed by an out-of-plane M–H loop (Fig. 1c), measured using a vibrating sample magnetometer (VSM). The three-step magnetization reversal process, from +400 to −400 mT, corresponds to the sequential switching of the free, coupling, and fixed layers. Switching of the free layer before the external field reaches zero indicates the presence of antiferromagnetic exchange coupling at zero field. Pillars with various diameters (100–5,000 nm) were fabricated via photolithography and electron-beam lithography. An oblique schematic of the patterned device and electrode configuration is shown in Fig. 1b. The reduced overlap between the top and bottom electrodes ensures efficient high-frequency (>1 GHz) pulse injection via a ground–signal–ground (GSG) probe.

*Contact author: jpwang@umn.edu

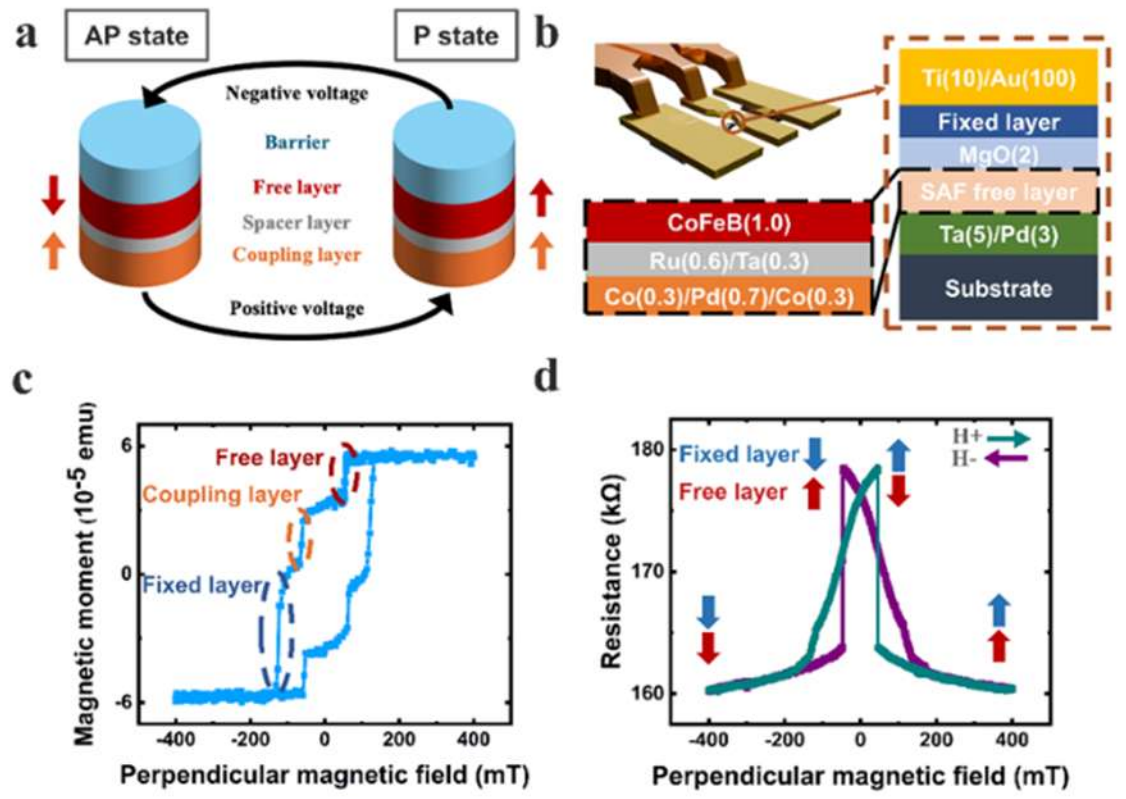

FIG. 1 | Device structure and magnetic properties of the EM-VCEC stack. (a) Schematic illustration of electron-mediated voltage-controlled exchange coupling (EM-VCEC)-induced switching. (b) Schematic of the pMTJ stack structure incorporating a SAF free layer. The top-left inset shows a schematic illustration of a patterned pillar with electrodes configured for GSG probe measurements. (c) Magnetization vs. perpendicular magnetic field loop of the stack after annealing. (d) Resistance vs. perpendicular magnetic field loop of a patterned pMTJ device (500 nm diameter).

Fig. 1d shows the resistance–field (R–H) loop of a 500 nm device, exhibiting distinct resistance states corresponding to different magnetization orientations of the free layer. The sharp low-to-high resistance transition as the magnetic field is reduced from 400 mT to 0 mT indicates the presence of both PMA and antiferromagnetic coupling within the pillar structure. The device exhibits a tunneling magnetoresistance (TMR) ratio of ~11% and a resistance–area (RA) product of $3.93 \times 10^4\ \Omega\cdot\mu m^2$, consistent with a 2 nm MgO tunnel barrier. Size-dependent resistance analysis (Supporting Material S1) verifies that the effective junction area closely matches the nominal device dimensions. Current–voltage (I–V) characteristics for devices with varying diameters are provided in Supporting Material S2.

Fig. 2a shows normalized minor hysteresis loops of a 500 nm device under various applied voltages. Before each measurement, the magnetization was saturated in the positive field direction to ensure a well-defined initial state, such that only the free layer reverses into the antiparallel (AP) configuration. A constant current was applied to monitor the resistance, and the average voltage between the AP and parallel (P) states was recorded for analysis.

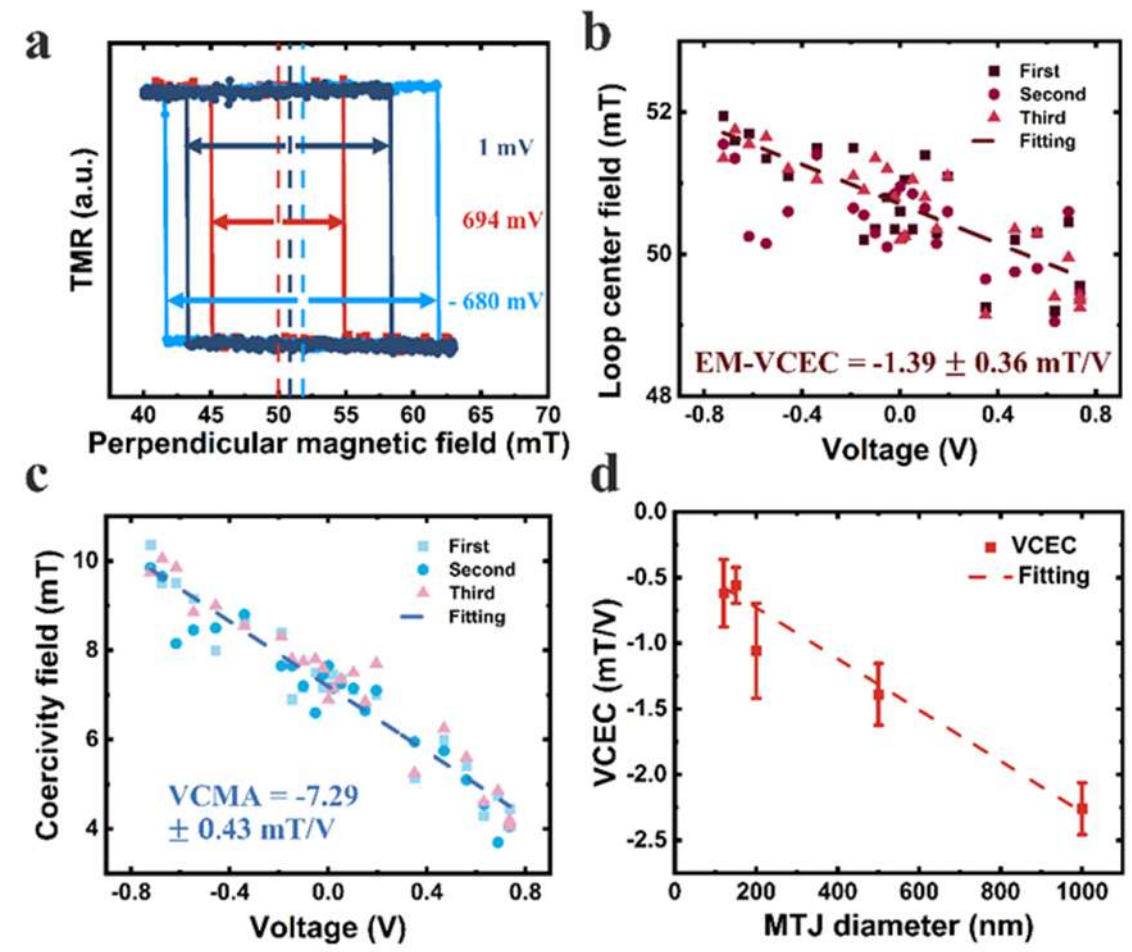

FIG. 2 | Voltage-induced effective fields from VCEC and VCMA. (a) Normalized minor hysteresis loops of a 500 nm MTJ under various applied voltages. Positive (negative) voltages narrow (broaden) the loop width due to VCMA, and shift the loop leftward (rightward) due to EM-VCEC. The voltage dependence of the (b) loop center field and (c) coercivity is extracted from minor loops. For each voltage, three loops are measured to reduce statistical uncertainty. Linear fits to the data yield the EM-VCEC and VCMA coefficients. (d) Extracted VCEC strength as a function of pillar diameter. Error bars represent fitting uncertainties.

Two voltage-dependent effects are anticipated: a loop shift due to VCEC and a width change due to VCMA. The VCMA effect alters the PMA of the free layer, reflected in the coercivity change. Under negative voltage, the minor loop widens due to enhanced interfacial anisotropy from electron depletion [5]. In contrast, VCEC manifests as a shift of the minor loop center, stemming from voltage-induced modulation of the exchange coupling field. The loop shifts left or right depending on voltage polarity. To quantify these effects, loop center field and coercivity were extracted from three repeated measurements per voltage. As shown in Fig. 2b,c, both exhibit near-linear voltage dependence, yielding VCEC and VCMA strengths of -1.39 ± 0.36 mT/V and -7.29 ± 0.43 mT/V, respectively.

Importantly, the observed loop shift is inconsistent with STT, as it (1) varies linearly with applied voltage, and (2) contradicts the STT polarity, where a positive current favors the AP state. This confirms a voltage origin, attributed to VCEC.

To further understand the nature of VCEC, we analyze its dependence on device geometry. As shown in Fig. 2d, the extracted VCEC strength decreases nearly linearly with pillar diameter, suggesting a size-dependent change likely caused by edge-related effects. This trend is consistent with a scaling relation in which the active central region follows S/D ~ d, where S is the pillar area, D the perimeter, and d the pillar diameter. In contrast, the VCMA strength shows no clear size dependence, possibly due to coercivity variations among nominally identical devices (see Supporting Information Section 4).

*Contact author: jpwang@umn.edu

The relatively weak VCEC compared to VCMA inhibits bipolar switching in our devices. Under positive voltage, VCMA lowers the energy barrier, enabling AP-to-P switching. In contrast, negative voltage raises the barrier between the P and AP states, effectively suppressing P-to-AP switching (see Supporting Material S3). Bipolar switching is expected when VCEC dominates over VCMA. The small VCEC strength may result from limited interlayer exchange coupling or non-uniform voltage distribution across the stack.

To investigate pMTJ response under fast voltage excitation, we used the circuit shown in Fig. 3a. A 5 nA DC bias current was applied through the bias-T's DC port to monitor resistance, while voltage pulses were injected through the AC port. The DC bias, far below the typical critical switching current, does not perturb the magnetization state. The test sequence is illustrated in Fig. 3b: a 1 mT out-of-plane field is first applied to initialize the AP state, then reduced to a bistable region where voltage pulses of varying widths and amplitudes are applied. After each pulse, the resistance is read to determine the magnetic state. Each condition was repeated 200 times to extract the switching probability.

Fig. 3c shows the minor loop of a 150 nm device, which has relatively larger coercivity and is selected to study the effect of the external magnetic field on switching. Magnetic fields ranging from 40 to 48 mT within the bistable region are applied. At each magnetic field, the voltage amplitude is fixed at 3.56 V, and the pulse width is gradually increased from 0.5 ns until the switching probability approaches approximately ~0.9. As shown in Fig. 3d, the switching probability increases with pulse width and is well-fitted by a sigmoid function. Switching speed generally increased with magnetic field strength from 40 to 48 mT, consistent with the role of a positive out-of-plane field in assisting AP-to-P switching. A broader range of magnetic fields (38–48 mT) was tested, with full data provided in Supporting Material S4. For clarity, only selected results are shown in Fig. 3d, all confirming that switching occurs under all conditions and accelerates with increasing field strength.

In a device influenced solely by VCMA, a positive voltage would contract the minor loop toward the center, and AP-to-P switching would occur only if the applied magnetic field exceeded the loop center field (illustrated in Fig. 3e). At lower fields, VCMA alone would favor the AP state during the initial pulse, suppressing AP-to-P switching. However, the observation of switching at relatively low fields (40 and 42 mT), where VCMA should be insufficient for switching, indicates the presence of an additional mechanism. This mechanism generates an effective field that assists switching and emerges within 2 ns of the voltage pulse, consistent with the characteristics of EM-VCEC.

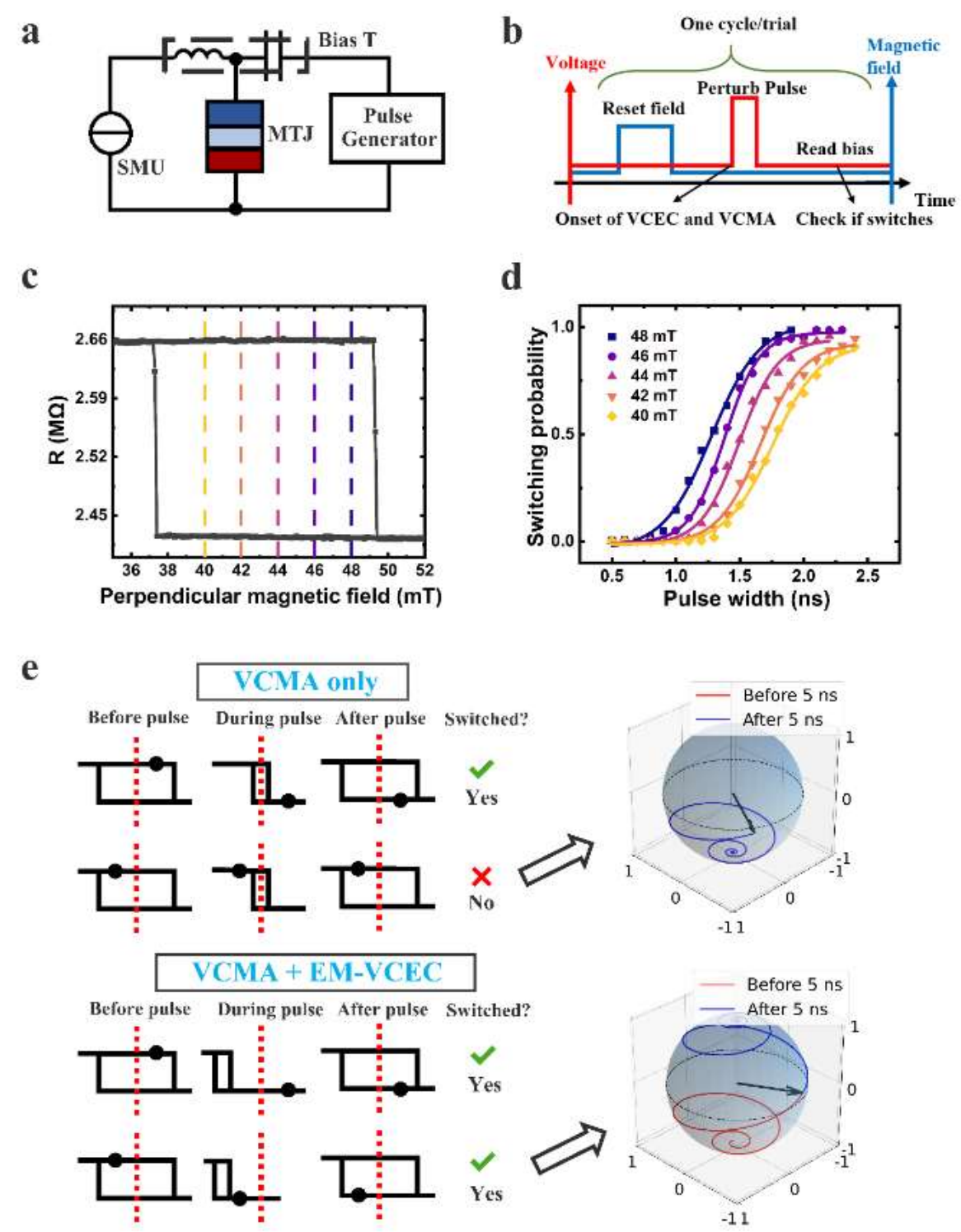

FIG. 3 | EM-VCEC and VCMA-induced switching using voltage pulses. (a) Schematic of the fast pulse switching test setup. (b) Test sequence for switching probability measurements. (c) Minor hysteresis loop of the 150 nm MTJ. The dashed line marks the external magnetic field used in the switching tests. (d) Switching probability as a function of pulse width, measured at a fixed pulse amplitude of 3.56 V under selected external magnetic fields. (e) Illustration of the evolution of the R–H minor loop under applied voltage. The black dot indicates the state of the free layer, which resides either on the left or right side of the loop center, depending on the external magnetic field. In contrast to conventional VCMA, EM-VCEC enables switching even when the free layer remains on the left side of the loop center. The trajectory from the macromagnetic simulation shows the detailed dynamic behavior of the switching with and without the EM-VCEC.

Because the switching process occurs on a nanosecond timescale, the hysteresis loop shift, as an equilibrium response, may not fully capture the underlying dynamics. To gain deeper insight, we perform macrospin simulations of the magnetization trajectory (see Supporting Material S5). In the absence of EM-VCEC, when only VCMA is applied, the PMA is reduced, and the magnetization begins to process from the −z direction toward the in-plane axis. However, if the external magnetic field favors the −z state, the magnetization fails to cross the equator and relaxes back after the voltage pulse ends. As shown in the top-right of Fig. 3e, because thermal fluctuations are not included, the trajectory after 5 ns exactly overlaps with that before 5 ns, indicating no net switching. In contrast, when EM-VCEC is activated and introduces an effective field along the +z direction, the magnetization crosses the equator during the pulse and stabilizes in the +z state afterward, thereby completing the switching. This result highlights the key role of EM-VCEC in overcoming the energy barrier imposed by the external field and enabling deterministic, fast magnetization reversal.

*Contact author: jpwang@umn.edu

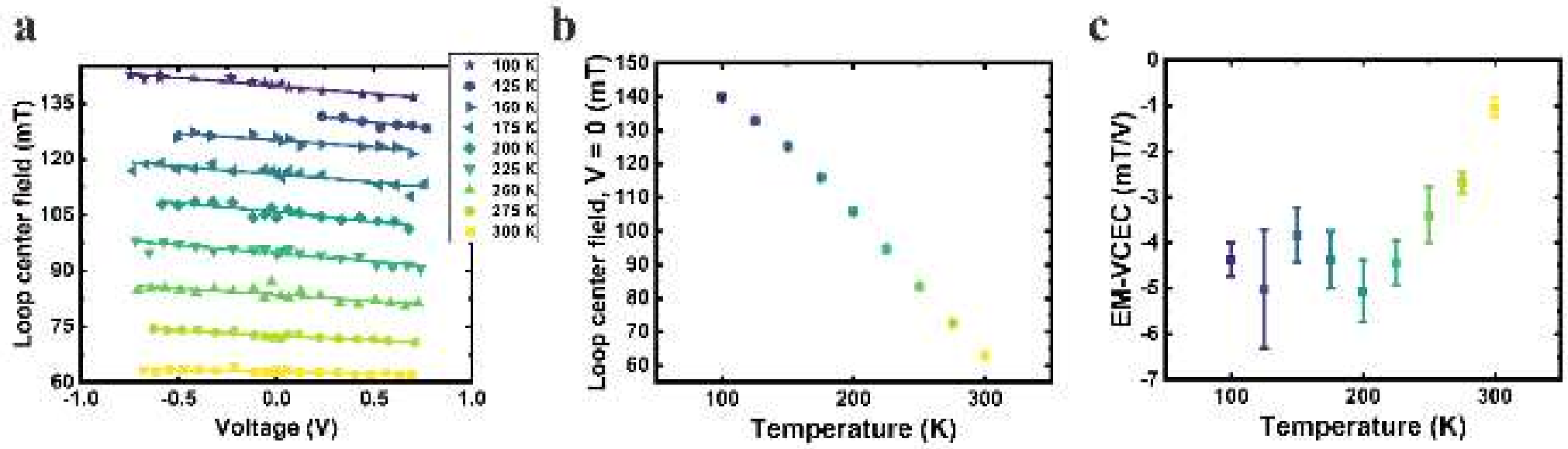

FIG. 4 | Temperature-dependent minor loop measurements. (a) Voltage-dependent loop center fields extracted at different temperatures. Linear fits are performed at each temperature, and the corresponding slopes and intercepts are plotted in (b) and (c), respectively. (b) Loop center field at 0 V as a function of temperature, reflecting the strength of antiferromagnetic (AFM) coupling. (c) EM-VCEC strength, represented by the extracted effective field (slope), as a function of temperature.

To further clarify the origin of the observed VCEC, we performed temperature-dependent minor loop measurements under various voltage conditions. A small magnetic field step size was used to accurately capture the field for both AP-to-P and P-to-AP transitions. From the loops, the loop center field was extracted, as shown in Fig. 4a. Across all measured temperatures, the loop center field exhibits a linear dependence on the applied voltage, confirming the presence of VCEC even at low temperatures. Linear fits were used to extract both the slope and the intercept (Figs. 4b and 4c), with the intercept representing the loop center field at zero voltage. Interestingly, the intercept increases monotonically as temperature decreases, indicating a strengthening of the antiferromagnetic (AFM) exchange coupling. This behavior is consistent with the temperature dependence of the RKKY interaction, which is known to enhance AFM coupling at lower temperatures, in agreement with previous reports [31].

The extracted slope, representing the EM-VCEC strength (the effective field per unit voltage), is shown in Fig. 4c. It increases as the temperature decreases from 300 K to 200 K, and then saturates, remaining nearly constant down to 100 K. This behavior rules out ion migration as a dominant mechanism, since ionic motion would be frozen at low temperatures, leading to a reduced effect. Instead, the temperature dependence supports an electron-mediated origin. The enhancement may arise from two factors: (1) stronger AFM exchange at lower temperatures, which amplifies the voltage-tunable exchange field, and (2) Fermi surface sharpening, which increases the system's voltage sensitivity. Further theoretical modeling is required to quantify the relative contributions of these two mechanisms.

In summary, we demonstrate the electronic origin of VCEC in perpendicular magnetic tunnel junctions (pMTJs) with a coupled free layer. Pulse-induced switching measurements confirm that the exchange field persists on the nanosecond timescale, providing direct experimental evidence that exchange coupling can be modulated electronically. This behavior clearly distinguishes EM-VCEC from magneto-ionic mechanisms, which rely on much slower ion migration. Furthermore, temperature-dependent measurements reveal that the EM-VCEC strength increases with decreasing temperature and eventually saturates at low temperatures, conclusively ruling out ion migration as the underlying mechanism. These findings establish EM-VCEC as a fast, deterministic, and energy-efficient mechanism for voltage-driven magnetization switching.

*Acknowledgments*—This work was partly supported by the Defense Advanced Research Projects Agency (DARPA) (Advanced MTJs for computation in and near random access memory) under Grant HR001117S0056-FP-042, the Global Research Collaboration (GRC) Logic and Memory program, sponsored by SRC and NSF ASCENT program TUNA: No. 2230963. The authors thank the useful discussion and support from Denis Tonini and Yu Han Huang, a visiting student from National Yang Ming Chiao Tung

*Contact author: jpwang@umn.edu

University, Taiwan. Parts of this work were carried out in the Characterization Facility, University of Minnesota, which receives partial support from the NSF through the MRSEC (Award Number DMR-2011401) and the NNCI (Award Number ECCS-2025124) programs. Portions of this work were conducted in the Minnesota Nano Center, which is supported by the National Science Foundation through the NNCI under Award Number ECCS-2025124. This research also is partially based upon work supported by the National Science Foundation Graduate Research Fellowship under Grant No. 2237827

*Contact author: jpwang@umn.edu

*Contact author: jpwang@umn.edu